# Reflected-light Phase Curves with PICASO: A Kepler-7b Case Study

Colin D. Hamill[1], Alexandria V. Johnson[1], Natasha Batalha[2], Rowan Nag[1], Peter Gao[3], Danica Adams[4], and Tiffany Kataria[5]

[1] Department of Earth, Atmospheric and Planetary Sciences, 550 Stadium Mall Drive, Purdue University, West Lafayette, IN 47907, USA; hamillc@purdue.edu  
[2] NASA Ames Research Center, MS 245-3, Moffett Field, CA 94035, USA  
[3] Carnegie Science Earth and Planets Laboratory, 5241 Broad Branch Road, NW, Washington, DC 20015, USA  
[4] Harvard University, Department of Earth and Planetary Sciences, 20 Oxford Street, Cambridge, MA 02138, USA  
[5] Jet Propulsion Laboratory, California Institute of Technology, 4800 Oak Grove Drive, Pasadena, CA 91109, USA

## Abstract

Examining reflected light from exoplanets aids in our understanding of the scattering properties of their atmospheres and will be a primary task of future flagship space- and ground-based telescopes. We introduce an enhanced capability of Planetary Intensity Code for Atmospheric Scattering Observations (PICASO), an open-source radiative transfer model used for exoplanet and brown dwarf atmospheres, to produce reflected light phase curves from three-dimensional atmospheric models. Since PICASO is coupled to the cloud code Virga, we produce phase curves for different cloud condensate species and varying sedimentation efficiencies ($f_{sed}$) and apply this new functionality to Kepler-7b, a hot Jupiter with phase curve measurements dominated by reflected starlight. We model three different cloud scenarios for Kepler-7b: $MgSiO_3$ clouds only, $Mg_2SiO_4$ clouds only, and $Mg_2SiO_4$, $Al_2O_3$, and $TiO_2$ clouds. All our Virga models reproduce the cloudy region west of the substellar point expected from previous studies, as well as clouds at high latitudes and near the eastern limb, which are primarily composed of magnesium silicates. $Al_2O_3$ and $TiO_2$ clouds dominate near the substellar point. We then compare our modeled reflected light phase curves to Kepler observations and find that models with all three cloud condensate species and low sedimentation efficiencies (0.03–0.1) match best, though our reflected light phase curves show intensities approximately one-third of those observed by Kepler. We conclude that a better understanding of zonal transport, cloud radiative feedback, and particle scattering properties is needed to further explain the differences between the modeled and observed reflected light fluxes.

*Unified Astronomy Thesaurus concepts:* Hot Jupiters (753); Atmospheric clouds (2180); Radiative transfer (1335); Exoplanet atmospheres (487); Extrasolar gaseous planets (2172)

## 1. Introduction

Phase curves are crucial for understanding the global climate and cloud coverage of exoplanets. For hot Jupiters, which are prime candidates for observation due to their size and orbital distance, visible and infrared phase curve observations from Hubble, TESS, and Spitzer have shown evidence of atmospheric dynamics, such as shifted hotspots due to strong equatorial winds and day/night temperature contrasts (e.g., H. A. Knutson et al. 2007; K. B. Stevenson et al. 2014, 2017; I. Wong et al. 2016, 2020, 2021; L. Kreidberg et al. 2018; J. Arcangeli et al. 2019; T. Daylan et al. 2021; A. Shporer et al. 2019; T. G. Beatty et al. 2020; T. Jansen & D. Kipping 2020; von Essen et al. 2020; E. M. May et al. 2021; G. Morello et al. 2023; Q. Changeat et al. 2024). Thermal phase curves have also begun to unveil the composition and thermal structure of warm sub-Neptune planets (e.g., GJ 1214b) using the James Webb Space Telescope (JWST; P. Gao et al. 2023; E. M.-R. Kempton et al. 2023). Phase curve observations from Kepler, which probe a planet's reflected light, have shown evidence of diverse albedos and brightness variations across hot Jupiter daysides (e.g., B.-O. Demory et al. 2013; L. J. Esteves et al. 2013, 2015; E. V. Quintana et al. 2013; A. Shporer et al. 2014; D. Angerhausen et al. 2015; A. Shporer & R. Hu 2015; B. Jackson et al. 2019). Some of these brightness variations are attributed to the presence of clouds.

Clouds are commonly inferred or detected from exoplanet observations, regardless of planetary size or temperature, and their presence complicates observations. The presence of clouds can mute atomic and molecular feature amplitudes in transmission spectra (e.g., D. Charbonneau et al. 2002; J. L. Bean et al. 2010; H. A. Knutson et al. 2014; L. Kreidberg et al. 2014; I. J. M. Crossfield 2015; D. K. Sing et al. 2016; N. P. Gibson et al. 2017; H. R. Wakeford et al. 2017; G. Bruno et al. 2018; P. C. Thao et al. 2020; J. Lustig-Yaeger et al. 2023), making the detection of atmospheric constituents difficult. They can also have direct, and drastically different, impacts on climate, depending on their chemical composition, particle size, and location within the atmosphere (M. S. Marley et al. 2013; C. Helling 2019; P. Gao et al. 2021). Observations from JWST are just beginning to detect the cloud composition of exoplanets (e.g., A. Dyrek et al. 2024; D. Grant et al. 2023), but degeneracies between cloud composition and particle size often prevent positive identification of a cloud species. Since visible light phase curves are heavily influenced by the presence of atmospheric particulates, phase curves dominated by reflected light from the host star have the potential to tell us a myriad of details about exoplanet clouds and the underlying atmospheric properties in the coming decades.

Kepler-7b (D. W. Latham et al. 2010) is a low-density (0.14 g cm$^{-3}$), short-period (4.88 days) planet with an equilibrium temperature ($T_{eq}$) of 1500–1600 K and a high albedo at visible wavelengths ($A_g \gtrsim 0.25$; B.-O. Demory et al. 2011, 2013; D. Kipping & G. Bakos 2011; D. Angerhausen et al. 2015; L. J. Esteves et al. 2015; K. Heng et al. 2021) orbiting a G-type star. Phase-resolved observations of Kepler-7b frequently show a westward shift in the phase maximum that has been attributed to a reflective cloud layer localized to the western half of the dayside hemisphere (D. Kipping & G. Bakos 2011; J. L. Coughlin & M. López-Morales 2012; B.-O. Demory et al. 2013; D. Angerhausen et al. 2015; L. J. Esteves et al. 2015). There have been multiple techniques used to simulate reflected-light phase variations for Kepler-7b and other hot Jupiters, all with their own underlying assumptions. Some of these techniques involved fitting cloud models to Kepler observations by treating the optical properties, thermal structure, or relative cloud abundances of the planet as free parameters (e.g., B.-O. Demory et al. 2013; K. Heng & B.-O. Demory 2013; R. Hu et al. 2015; A. G. Muñoz & Isaak 2015; M. Oreshenko et al. 2016). Another approach was to model $Mg_2SiO_4$, $MgSiO_3$, and Fe condensation, using the cloud model Virga (A. S. Ackerman & M. S. Marley 2001; N. Batalha et al. 2020a; C. M. Rooney et al. 2022), at specified longitudes using a grid of one-dimensional pressure/temperature profiles with calculated cloud opacity (M. W. Webber et al. 2015).

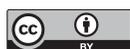

Three-dimensional (3D) general circulation models (GCMs) are often used to interpret phase curve observations, due to their ability to simulate spatially heterogeneous atmospheric properties, such as temperature, chemical composition, wind profiles, and cloud distributions (e.g., I. Dobbs-Dixon & E. Agol 2013; N. K. Lewis et al. 2014; R. T. Zellem et al. 2014; T. Kataria et al. 2015, 2016; D. S. Amundsen et al. 2016; I. Wong et al. 2016; N. K. Lewis et al. 2017; K. B. Stevenson et al. 2017; J. M. Mendonça et al. 2018; V. Parmentier et al. 2018; J. Arcangeli et al. 2019; E. Flowers et al. 2019; M. E. Steinrueck et al. 2019; O. Venot et al. 2020; C. K. Harada et al. 2021; V. Parmentier et al. 2021; M. T. Roman et al. 2021; N. Robbins-Blanch et al. 2022; J. W. Skinner & J. Y.-K. Cho 2022). Though not applied specifically to Kepler-7b, V. Parmentier et al. (2016) modeled equilibrium cloud condensation using 3D GCMs of generic hot Jupiter atmospheres and produced reflected and emitted theoretical phase curves. They found that a westward shift in the reflected phase curve exists for cooler ($T_{eq} \lesssim 1900$ K) hot Jupiters, due to the lower temperatures and subsequent cloud formation west of the substellar point. M. Roman & E. Rauscher (2017) used 3D GCMs of Kepler-7b with cloud radiative feedback to model infrared phase curves of Kepler-7b, though they prescribe fixed cloud locations, including vertical extent, from the best-fitting models of B.-O. Demory et al. (2013) and A. G. Muñoz & K. G. Isaak (2015). They found that inhomogeneous cloud coverage plays an important role in regulating equatorial jet strength and dayside/nightside temperature contrasts, and that clouds likely form at high latitudes near the western terminator.

A variety of models have been used to investigate the effects of clouds in exoplanet atmospheres. Multiple cloud models used today (e.g., *EddySed*, *Exo-REM*, and Virga) are based off of the equilibrium model by A. S. Ackerman & M. S. Marley (2001), which balances particle sedimentation with lofting from eddy diffusion to compute condensate particle size distributions and cloud vertical extent (e.g., M. S. Marley et al. 2010; C. V. Morley et al. 2012, 2014; J.-L. Baudino et al. 2015, 2017; B. Charnay et al. 2018; D. A. Christie et al. 2021; C. M. Rooney et al. 2022). *CLIMA* goes a step further by coupling *EddySed* clouds to a 1D radiative-convective climate model for terrestrial planets with patchy clouds (J. D. Windsor et al. 2023), while *YunMa* uses the cloud-modeling approach from A. S. Ackerman & M. S. Marley (2001) within a retrieval package for transit spectra (S. Ma et al. 2023). More complicated tools (e.g., *CARMA* and *DRIFT*) model aerosol formation through microphysical processes including nucleation, evaporation, condensation, and coagulation (e.g., C. Helling et al. 2008; C. Helling & A. Fomins 2013; P. Gao et al. 2018, 2020; Y. Kawashima & M. Ikoma 2018; D. Powell et al. 2019). Other exoplanet cloud models are instead focused on better characterizing the optical properties of nonspherical cloud particles, as a variety of solid-phase aerosols likely form aggregates or crystals instead of spheres (R. Tazaki & H. Tanaka 2018; K. Ohno et al. 2020; D. Samra et al. 2020, 2022; K. L. Chubb et al. 2024).

In this study, we introduce a new function to the Planetary Intensity Code for Atmospheric Scattering Observations (PICASO) for computing reflected-light phase curves from GCM output. Using the temperature/pressure and eddy diffusion coefficient values from the GCM of Kepler-7b (presented in D. J. Adams et al. 2022) as input, we post-process chemistry within PICASO and utilize the Virga cloud model, which is coupled to PICASO, to simulate the vertical extent, particle sizes, and cloud opacities of multiple cloud condensate species. We benchmark our methodology by comparing PICASO output to Kepler phase curve observations of Kepler-7b. As PICASO and Virga are also open source, the new phase curve function described here will be available to the public and applicable to a wide range of exoplanets.

In Section 2, we describe Virga and PICASO in more detail and outline our updates to PICASO. In Section 3, we present our Virga-computed cloud profiles and our synthetic cloudy and cloudless reflected-light phase curves of Kepler-7b. We also compare our results to Kepler phase curve data from B.-O. Demory et al. (2013). Our conclusions are presented in Section 4.

## 2. Models

### 2.1. Virga

Virga, an open-source cloud model developed by N. Batalha et al. (2020b), is used to compute the optical depths, single-scattering albedo, and asymmetry parameters for phase-equilibrium condensate clouds as a function of atmospheric pressure and wavelength. Virga is based on *EddySed* developed by A. S. Ackerman & M. S. Marley (2001), which has been used in a variety of cases to model the 1D or 3D effects that equilibrium cloud decks have on the scattering properties of substellar atmospheres (e.g., J. J. Fortney et al. 2006; C. V. Morley et al. 2012, 2013, 2015; D. J. Adams et al. 2022; N. Robbins-Blanch et al. 2022).

Virga uses the sedimentation efficiency ($f_{sed}$), a tunable parameter, to control the vertical extent of clouds—and by extension, the particle size distribution. A. S. Ackerman & M. S. Marley (2001) introduced this parameter and defined it as the ratio of the mass-weighted droplet sedimentation velocity to $w_*$, the convective velocity scale. As such, the vertical extent of a cloud deck is dependent on the balance between the downward $f_{sed} \, w_*$ term and the upward convective term, parameterized by the vertical eddy diffusion coefficient ($K_{zz}$). In our Virga models, we calculate a global average $K_{zz}$ profile from our Kepler-7b GCM and use this profile in every 64 × 128 lat–lon grid point. We consider a global average $K_{zz}$ profile to reduce numerical instability, which occurs when using the individual grid point $K_{zz}$ profiles because GCMs treat both upward and downward mixing, whereas Virga only treats upward mixing. For example, while rising air on the dayside could mix cloud particles upward, the downward branch on the nightside would instead concentrate cloud particles near the cloud base, acting in the same direction as sedimentation rather than against it. Global $K_{ZZ}$ profiles are often used in global circulation models because mass transport on tidally locked hot Jupiters is inherently global and may not be purely diffusional (X. Zhang & A. P. Showman 2018, D. J. Adams et al. 2022). The $K_{zz}$ minimum is set to $10^5 \, \mathrm{cm^2 \, s^{-1}}$, which is the default minimum set by Virga. The cloud distribution at equilibrium can then be described by

$$-K_{zz}\frac{\partial q_t}{\partial z} - f_{\mathrm{sed}} w_* q_c = 0, \quad (1)$$

where $q_t$ is the total mixing ratio of condensate and vapor, $q_c$ is the mixing ratio of condensates, and $z$ refers to atmospheric height. This equation is computed independently for each cloud condensate species and at each altitude level. Previous studies have suggested $f_{\mathrm{sed}}$ can range from 0.01 for warm mini-Neptunes to greater than 5 for giant planets and brown dwarfs (D. Saumon & M. S. Marley 2008; C. V. Morley et al. 2013, 2015; A. J. Skemer et al. 2016; R. J. MacDonald et al. 2018). In this study, we choose an $f_{\mathrm{sed}}$ range of 0.03–1.0 based on previous modeling efforts for matching hot Jupiter observations using Virga (e.g., M. W. Webber et al. 2015; D. A. Christie et al. 2021; D. J. Adams et al. 2022; N. Robbins-Blanch et al. 2022). The resulting cloud particles are assumed to be spherical, and therefore appropriate for the application of the Mie approximation for computing their optical properties. We use the appropriate condensate refractive indices from N. Batalha & M. Marley (2020).

Virga is currently capable of modeling thirteen different condensate species, namely $Al_2O_3$, $CH_4$, Cr, Fe, $H_2O$, KCl, $Mg_2SiO_4$, $MgSiO_3$, MnS, $NH_3$, $Na_2S$, $TiO_2$, and ZnS (N. Batalha et al. 2020a). Microphysical models that consider the impact of low cloud nucleation rates due to large energy barriers predict that only three of these condensates are likely to form in significant amounts in the observable part of Kepler-7b's atmosphere, $Mg_2SiO_4$, $Al_2O_3$, and $TiO_2$ (D. Powell et al. 2019; P. Gao et al. 2020; D. J. Adams et al. 2022). This three-cloud scenario assumes all SiO in the atmosphere is incorporated into $Mg_2SiO_4$ rather than $SiO_2$ or $MgSiO_3$, and similarly that all Mg will be consumed by $Mg_2SiO_4$ rather than $MgSiO_3$. The other condensates were predicted to not significantly contribute to the total cloud mass, due to a combination of limited elemental abundances and low nucleation rates, and thus we do not treat them here. We also model the cloud scenario where only $Mg_2SiO_4$ formation is considered, as it is predicted by microphysical models to dominate the aerosol optical depth of hot Jupiters with equilibrium temperatures like that of Kepler-7b (P. Gao et al. 2020). However, other hot Jupiter cloud models (e.g., M. W. Webber et al. 2015; V. Parmentier et al. 2016) have modeled $MgSiO_3$ as the dominant magnesium silicate cloud species, and thus we also consider a scenario in which only $MgSiO_3$ formation occurs.

## 2.2. PICASO

PICASO is an open-source radiative transfer model that computes the 1D or 3D thermal emission, transmission, and/or reflected-light spectra of exoplanets (N. E. Batalha et al. 2019; N. Batalha & Rooney 2020). In this study, we are considering reflected spectra only. PICASO is based on the radiative transfer equation (R. M. Goody & Y. L. Yung 1989; N. E. Batalha et al. 2019; D. J. Adams et al. 2022),

$$I(\tau_i, \mu) = I(\tau_{i+1}, \mu)\exp\left(\frac{\delta\tau_i}{\mu}\right) - \int_0^{\delta\tau_i} S(\tau', \mu)\exp\left(\frac{-\tau}{\mu}\right)d\tau'/\mu, \quad (2)$$

where $I(\tau_i, \mu)$ is the azimuthally averaged intensity emerging from the top of an atmospheric layer, $i$, with opacity, $\tau$, and outgoing angle, $\mu$. $I(\tau_{i+1}, \mu)\exp\left(\frac{\delta\tau_i}{\mu}\right)$ is the incident intensity on the lower boundary of the layer attenuated by the optical depth within the layer, $\delta\tau$, and $S(\tau', \mu)$ is the source function integrated over all layers. The source function is comprised of a single-scattering term and a multiple-scattering term given by

$$S(\tau', \mu) = \frac{\omega}{4\pi}F_0 P_{\mathrm{single}}(\mu, -\mu_0)\exp\left(-\frac{\tau'}{\mu_s}\right) + \frac{\omega}{2}\int_{-1}^{1} I(\tau', \mu)P_{\mathrm{multi}}(\mu', \mu)d\mu' \quad (3)$$

where $P_{\mathrm{single}}(\mu, -\mu_0)$ and $P_{\mathrm{multi}}(\mu', \mu)$ are the phase functions computed from single-scattered and multiple-scattered radiation, respectively. $\omega$ is the single-scattering albedo, and $F_0$ is the direct stellar radiation upon the atmosphere. The code has partial heritage from a Fortran-based albedo spectra model used for Solar System and exoplanet studies (C. P. McKay et al. 1989; O. B. Toon et al. 1989; M. S. Marley & C. P. McKay 1999; M. S. Marley et al. 1999; J. J. Fortney et al. 2005, 2006, 2008, 2010; K. L. Cahoy et al. 2010; T. Kataria et al. 2013, 2014, 2015, 2016; C. V. Morley et al. 2013, 2015; V. Parmentier et al. 2013, 2016), and the routine of O. B. Toon et al. (1989) still remains the default radiative transfer scheme in PICASO.

Properties of the planet/star system (i.e., orbital semimajor axis, phase angle, the planetary mass and radius, stellar mass, radius, flux, and metallicity) along with a pressure/temperature profile of the atmosphere are basic inputs for relative-flux reflected-light calculations. Before computing relative flux, PICASO computes the planet's geometric albedo, the reflected planet flux at full phase divided by the flux of a perfect Lambert disk (K. L. Cahoy et al. 2010; N. E. Batalha et al. 2019),

$$A_g(\lambda) = \frac{F_P(\alpha = 0°, \lambda)}{F_{0,L}(\lambda)}, \quad (4)$$

where $\alpha$ is the phase angle and $\lambda$ is wavelength. The ratio of planetary to stellar flux can then be calculated by

$$\frac{F_P(\lambda, \alpha)}{F_\odot(\lambda)} = A_g(\lambda)\left(\frac{R_P}{d}\right)^2 \Phi(\lambda, \alpha), \quad (5)$$

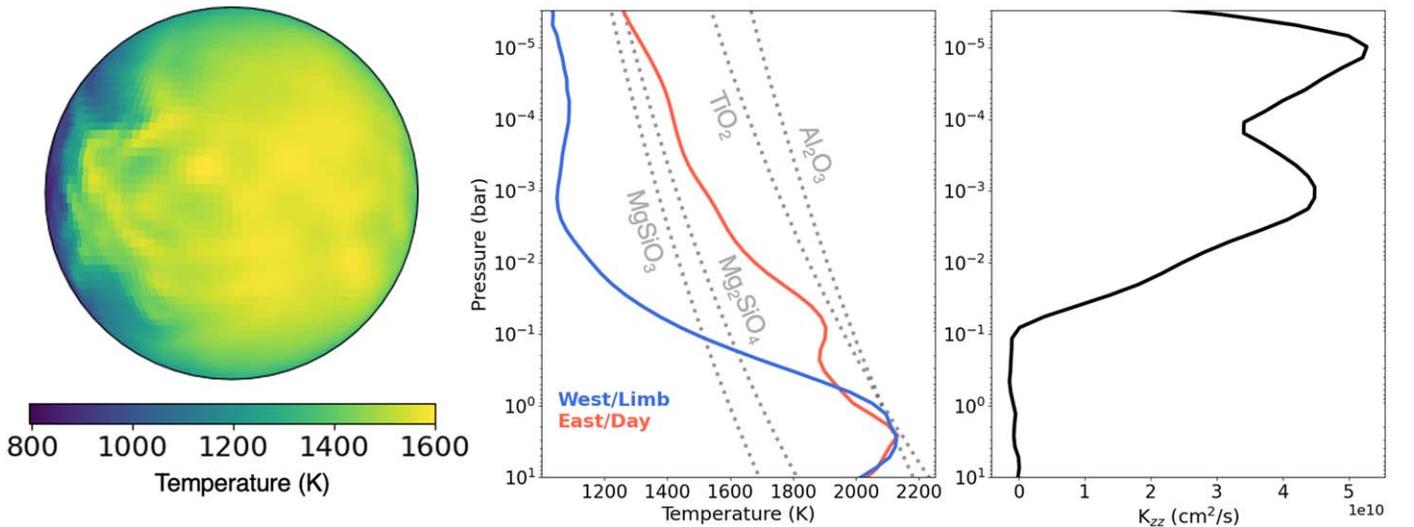

**Figure 1.** Left: temperature map of Kepler-7b's dayside hemisphere at 1 mbar. Middle: averaged temperature profiles of Kepler-7b for the western limb (blue) and hot dayside (red) regions, with the condensation curves (gray dashed) of the clouds ($MgSiO_3$, $Mg_2SiO_4$, $Al_2O_3$, and $TiO_2$) modeled in this study. Right: globally averaged $K_{zz}$ profile of the atmosphere. Low/negative $K_{zz}$ values at pressures >0.1 bar are set to the $K_{zz}$ minimum ($10^5$ cm$^2$ s$^{-1}$) before calculating Virga clouds. The SPARC/MITgcm data and condensation curves are from D. J. Adams et al. (2022).

**Table 1**
Properties of the Kepler-7b System

| Parameter | Value |
|---|---|
| $M_P$ ($M_J$) | 0.44 |
| $R_P$ ($R_J$) | 1.62 |
| $T_*$ (K) | 5933 |
| $R_*$ ($R_\odot$) | 1.97 |
| $a$ (au) | 0.06 |
| [Fe/H] | 0.11 |
| log $g$ (cgs) | 3.98 |

**Note.** From L. J. Esteves et al. (2015).

where $F_P(\lambda, \alpha)$ is the monochromatic flux of the planet, $F_\odot(\lambda)$ is the monochromatic flux of the star, $R_P$ is the stellar radius, $d$ is the planet–star distance, and $\Phi(\lambda, \alpha)$ is the planetary phase curve at phase angle $\alpha$. This equation does not consider thermal emission.

The Kepler-7b parameters used in this study are given in Table 1. We use the pressure/temperature (and globally averaged $K_{zz}$, for Virga) profiles of the Kepler-7b SPARC/MITgcm (64 × 128 lat–lon grid points) from D. J. Adams et al. (2022) as input (Figure 1). PICASO then computes the chemical composition of the atmosphere at every point in the pressure/temperature profile using a program adopted from S. Gordon & B. J. McBride (1994) by C. Visscher & J. I. Moses (2011), which assumes thermochemical equilibrium using the Sonora Bobcat model grid (M. S. Marley et al. 2021), which simulates the behavior of Fe-, Mg-, and Si-bearing gasses using the approach of C. Visscher et al. (2010). While recent 1D studies with PICASO have shown the importance of including disequilibrium chemistry for giant planets with strong vertical mixing (S. Mukherjee et al. 2022, 2023, 2024), PICASO does not yet have the capability to compute 3D disequilibrium chemistry. PICASO uses gas opacities from the Resampled Opacity Database for PICASO v2 (N. Batalha et al. 2020a), which includes opacities from 37 molecules and compounds. For the wavelength range considered in this study (350–950 nm), the main sources of gas opacity come from $H_2O$, Na, and K, with line lists from F. Allard et al. (2007), N. F. Allard et al. (2007, 2016, 2019), T. Ryabchikova et al. (2015), and O. L. Polyansky et al. (2018). If clouds can form in the atmosphere, then PICASO will use the pressure- and wavelength-dependent scattering parameters computed by Virga for said clouds. By default, PICASO uses Two-Term Henyey Greenstein phase functions for cloud scattering, which are weighted with the Rayleigh scattering phase function based on the relative contribution of each scattering component, to simulate scattering of the total atmosphere.

PICASO has the capability of simulating 1D (N. E. Batalha et al. 2019; J. Fraine et al. 2021; S.-Y. Tang et al. 2021) and 3D reflected and thermal emission spectra (D. J. Adams et al. 2022; N. Batalha et al. 2024), where phase-dependent spectra are calculated by using the Chebyshev–Gauss integration method to integrate over all gridded dayside intensities and compute geometric albedos, which was adapted from the methodology developed in K. L. Cahoy et al. (2010) to compute phase-dependent reflected-light fluxes. This method works by gridding the atmosphere into an array of plane–parallel facets, each with their own incident and observed angles. The chemical abundances and temperature profiles for this grid are computed using bilinear interpolation. The reflected intensity of the atmosphere is then computed by weighing each facet by its lat/lon location on the hemisphere visible to the observer, such that facets toward the middle of the disk are weighed more heavily than facets near the limbs. Facet longitude is weighted with Gauss–Legendre quadrature, and facet latitude is weighted with Chebyshev polynomials given by

$$w_i = \frac{\pi}{n+1}\sin^2\left(\frac{i}{n+1}\pi\right), \quad (6)$$

where $w_i$ is the Chebyshev weighting factor for Chebyshev point $i$, and $n$ is the number of Chebyshev points (20 in this study) along the latitudinal grid (M. Abramowitz & I. A. Stegun 1972).

D. J. Adams et al. (2022) used GCM simulations of six hot Jupiters as inputs to PICASO, including the Kepler-7b model used in this study (Figure 1), and found that these hot Jupiters have a diversity of albedos with no single explanation for their observable differences. N. Robbins-Blanch et al. (2022) updated PICASO to compute full-orbit (∼0°–360°) thermal emission phase curves using a 3D model atmosphere. N. Robbins-Blanch et al. (2022) achieved this by rotating a GCM grid such that the correct hemisphere is facing the observer at every discrete phase angle calculated by PICASO. They found that a phase curve produced using 15 discrete grid points and a 10 × 10 lat–lon grid to represent a hemisphere is enough to characterize the shape of the phase curve without adding unnecessary computation time. N. Robbins-Blanch et al. (2022) applied PICASO's new capability to the hot Jupiter WASP-43b and found good agreement between their models and WASP-43b observations from Hubble Space Telescope Wide-Field Camera 3 (WFC3) and at Spitzer's 3.6 and 4.5 m bands.

### 2.2.1. Reflected-light Phase Curves in PICASO

Our study makes a few significant changes to the thermal phase curve routine created by N. Robbins-Blanch et al. (2022) in order to include reflected-light phase curve capabilities. The main change is related to the geometry of how reflected-light calculations must be considered. In the thermal emission regime, the entire hemisphere seen by the observer emits infrared radiation. For a reflected-light calculation, however, the reflecting portion of the planet is constantly coming in or out of view, depending on how much of the dayside is visible to the observer. Our updated phase curve routine now tracks the waxing and waning of the reflecting dayside hemisphere as the planet travels through its orbit, and considers only reflected light from the dayside hemisphere (Figure 2).

PICASO is coupled to Virga to easily integrate post-processed clouds into PICASO's scattering approximations (N. Batalha 2020). The flow of information from Virga to PICASO in this study works as follows:

(1) Virga uses Equation (1) to create particle size distributions at each pressure level for the cloud condensates of interest.
(2) Virga uses the refractive indices and Mie parameters from PyMieScatt (N. Batalha & M. Marley 2020) to output the three optical constants needed by PICASO for each cloud condensate: single-scattering albedo ($\omega$), asymmetry parameter ($g$), and optical depth (opd) at each pressure level within the GCM.
(3) The full-resolution (64 lat × 128 lon) array of $\omega$, $g$, and opd, with coordinates of longitude, latitude, pressure, and wavenumber, is then input into PICASO.
(4) The full-resolution array is downsampled to a 20 × 20 grid for the dayside hemisphere seen by the observer for each discrete phase angle considered by PICASO.
(5) PICASO calculates the reflected intensity for each grid point and uses the Chebyshev–Gauss integration method to calculate total dayside reflected intensity.

The dayside portion seen by the observer is represented by a 20 × 20 lat–lon grid for this study at all phase angles in order to preserve the Chebyshev–Gauss integration method used throughout PICASO, which uses an $n \times m$ grid for all phases. Therefore, to update the phase curve routine without altering any of the underlying intensity integration methods, our grid resolution fluctuates throughout with respect to phase angle. In other words, the spatial resolution increases as phase approaches ±180° and decreases as phase approaches 0°. We find that a 20 × 20 spatial resolution offers the best balance between precision and computation time for our reflected-light calculations (see Appendix). A 20 × 20 grid spans around 167° of longitude for a phase of 0° and around 11° for a phase of ±168°, the largest absolute phase angles considered in our phase curve computation (see Section 3.1). The longitudinal and latitudinal grid points are set by the Chebyshev–Gauss integration method outlined in N. Robbins-Blanch et al. (2022). Both the GCM (Figures 1 and 2) and the Virga clouds (Figure 3) are independently rotated and regridded using the same method. Our techniques for computing reflected phase curves differ from previous studies focused on Kepler-7b (Section 1) because all the variables we use to compute reflected-light phase curves (besides sedimentation efficiency; see Section 2.1) are calculated a priori using a robust GCM, cloud model, and radiative transfer scheme. We posted our changes to PICASO to GitHub,[6] along with a tutorial,[7] where they are publicly available to the community.

## 3. Results and Discussion

### 3.1. Clouds on Kepler-7b

To simulate the impact that clouds have on Kepler-7b's albedo, we first calculate the vertical extent and scattering properties of multiple cloud condensate species as one-dimensional columns for every grid point in our 64 × 128 GCM. Figure 3 shows the optical depth maps at a wavelength of 604 nm and a pressure of 1 mbar for the three-cloud scenario ($Mg_2SiO_4$, $Al_2O_3$, and $TiO_2$) with $f_{sed} = 0.03$. We present the optical depth map of this cloud scenario because it provides the best fit to the observational data (see Section 3.2.1). We choose a wavelength of 604 nm for illustration because it is near the peak of the Kepler response function (KRF). Figure 3 shows that Kepler-7b's optically thick clouds are primarily located on the western side of the substellar point, which matches the results of previous Kepler-7b phase curve analyses (B.-O. Demory et al. 2013; L. J. Esteves et al. 2013; R. Hu et al. 2015; A. G. Muñoz & Isaak 2015; M. W. Webber et al. 2015). As predicted by some of the theoretical models before this study (V. Parmentier et al. 2016; M. Roman & E. Rauscher 2017), our Virga results suggest optically thick clouds may form at high latitudes. Around the substellar point and to the east of it, temperatures increase substantially such that cloud formation and opacity is much lower compared to the limb regions. The eight panels on the right side of Figure 3 show an example of how PICASO regrids the Virga cloud scattering map in the same manner outlined in Section 2.2.1. Although this regridding changes the cloud optical depth values slightly with respect to phase angle, it does not significantly impact the reflected-light curves. Regions west of the substellar point are preferentially viewed after secondary eclipse (positive phase angles), while regions east of the substellar point are preferentially viewed before secondary eclipse (negative phase angles). The first and last phases are both 0°, so there are seven unique phase angles considered in the example shown in Figure 3. For the phase curves shown in Section 3.2, 15 unique phase angles are used in order to provide better phase resolution.

---

[6] Code and documentation available at https://natashabatalha.github.io/picaso/.
[7] https://natashabatalha.github.io/picaso/notebooks/9g_ReflectedPhaseCurve.html

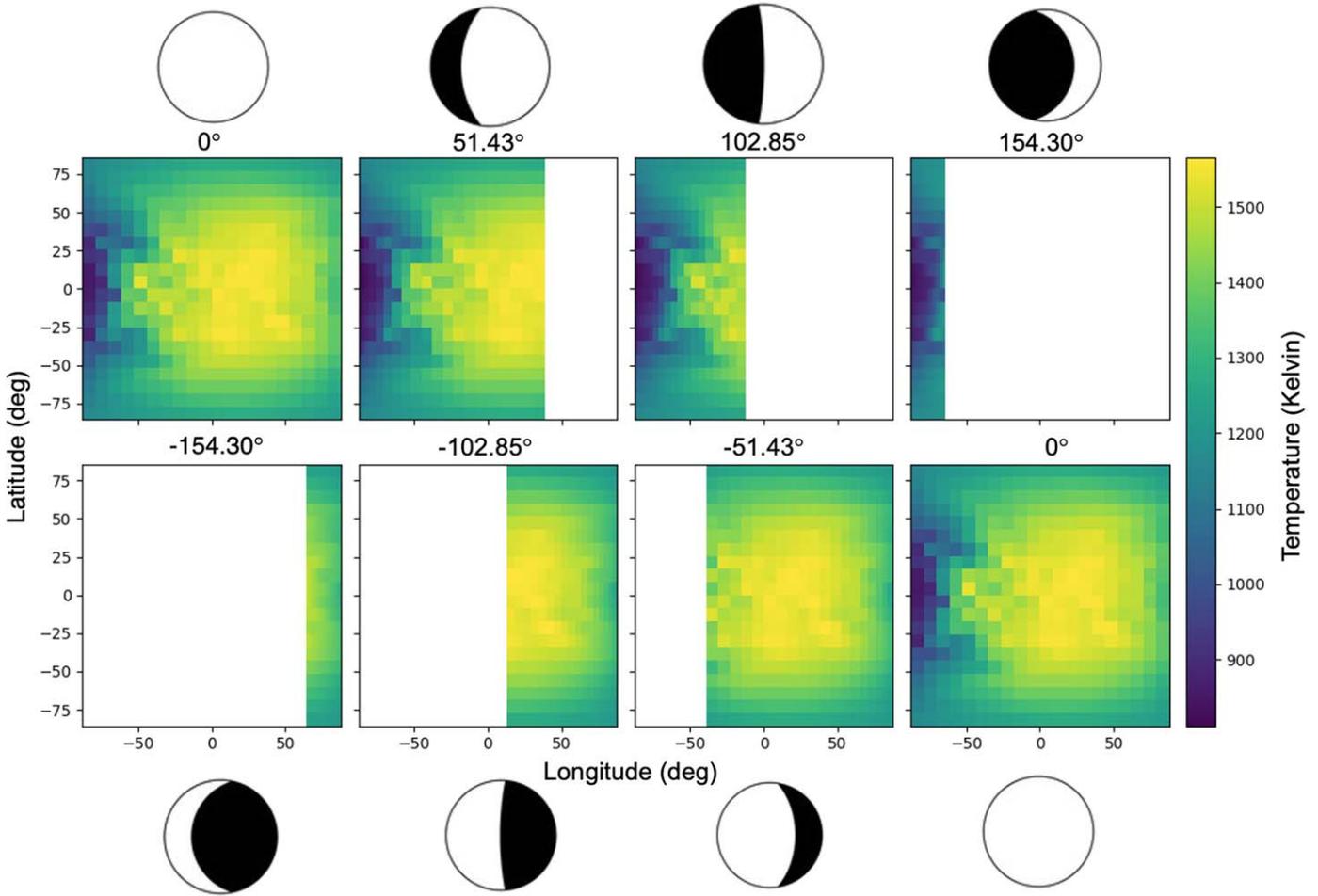

**Figure 2.** Temperature maps of the dayside hemisphere at 1 mbar illustrating how PICASO regrids and rotates the GCM for eight discrete planetary phases with 20 × 20 spatial grid resolution. The substellar point is located at the center of each map (0° latitude, 0° longitude), meaning these maps show the dayside hemisphere of Kepler-7b. Only colored regions contribute to reflected-light intensity. A phase angle of 0° corresponds to the secondary eclipse where the entire dayside hemisphere is visible. A phase angle of 180° (not shown) corresponds to the primary eclipse (transit). The planetary phase diagrams, which are included for reference, show the hemisphere seen by the observer as the planet orbits and both the dayside (white) and nightside (black) regions.

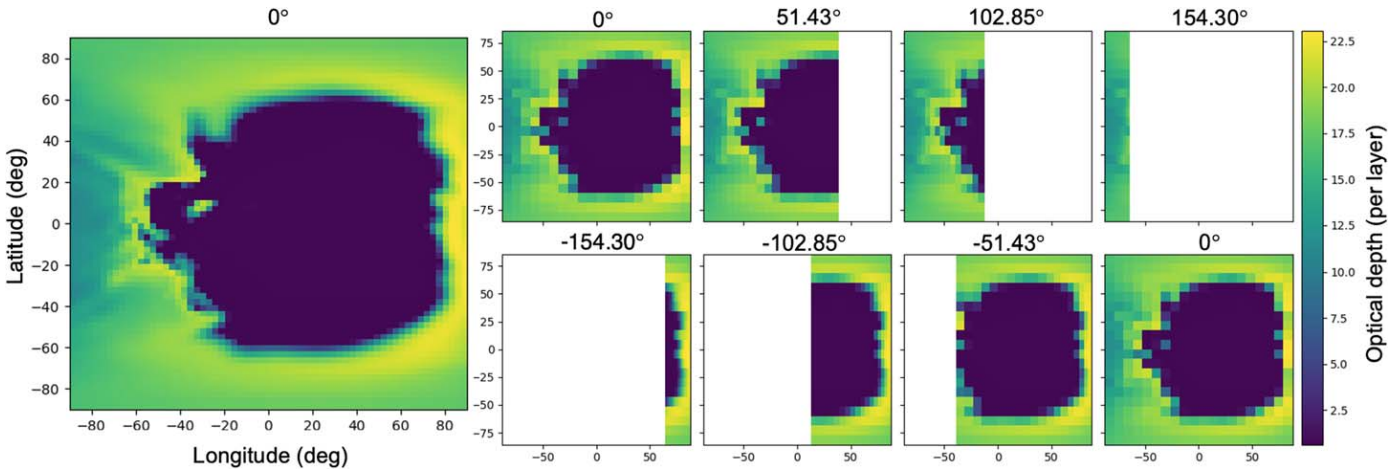

**Figure 3.** Left: full-resolution optical depth map of the dayside hemisphere at 1 mbar and a wavelength of 604 nm. Right: eight corresponding 20 × 20 grid phase maps used to compute reflected-light spectra. The optical depth maps here are generated with Virga considering $Mg_2SiO_4$, $Al_2O_3$, and $TiO_2$ as cloud condensates with $f_{sed} = 0.03$. The substellar point is located at the center of each map (0° latitude, 0° longitude).

Figure 4 maps the optical depth per layer as a function of longitude and pressure for every cloudy scenario considered in this study at a latitude close to the equator (1°.4N). Optically thick clouds (yellow, green) dominate the western terminator for all cloudy models shown, and the optical depth increases as sedimentation efficiency decreases. Cloud formation also occurs near the eastern terminator for all the cloudy cases shown. While the cloud scenarios with only $MgSiO_3$ or $Mg_2SiO_4$ formation have very similar cloud locations and optical depths near the equator, including an entirely cloudless region from $\sim -10°$ to $55°$ longitudes, the three-cloud scenario has full cloud coverage over the dayside hemisphere. This increase in cloud coverage is due to the refractory condensates, $Al_2O_3$ and $TiO_2$, forming at higher temperatures, albeit at much lower optical depths than regions where $Mg_2SiO_4$ formation is present. There is no cloud formation in any of the Virga models at pressures between $\sim 10$ and $10^{-1}$ bars, due to a strong inversion layer present in the Kepler-7b GCM (Figure 1).

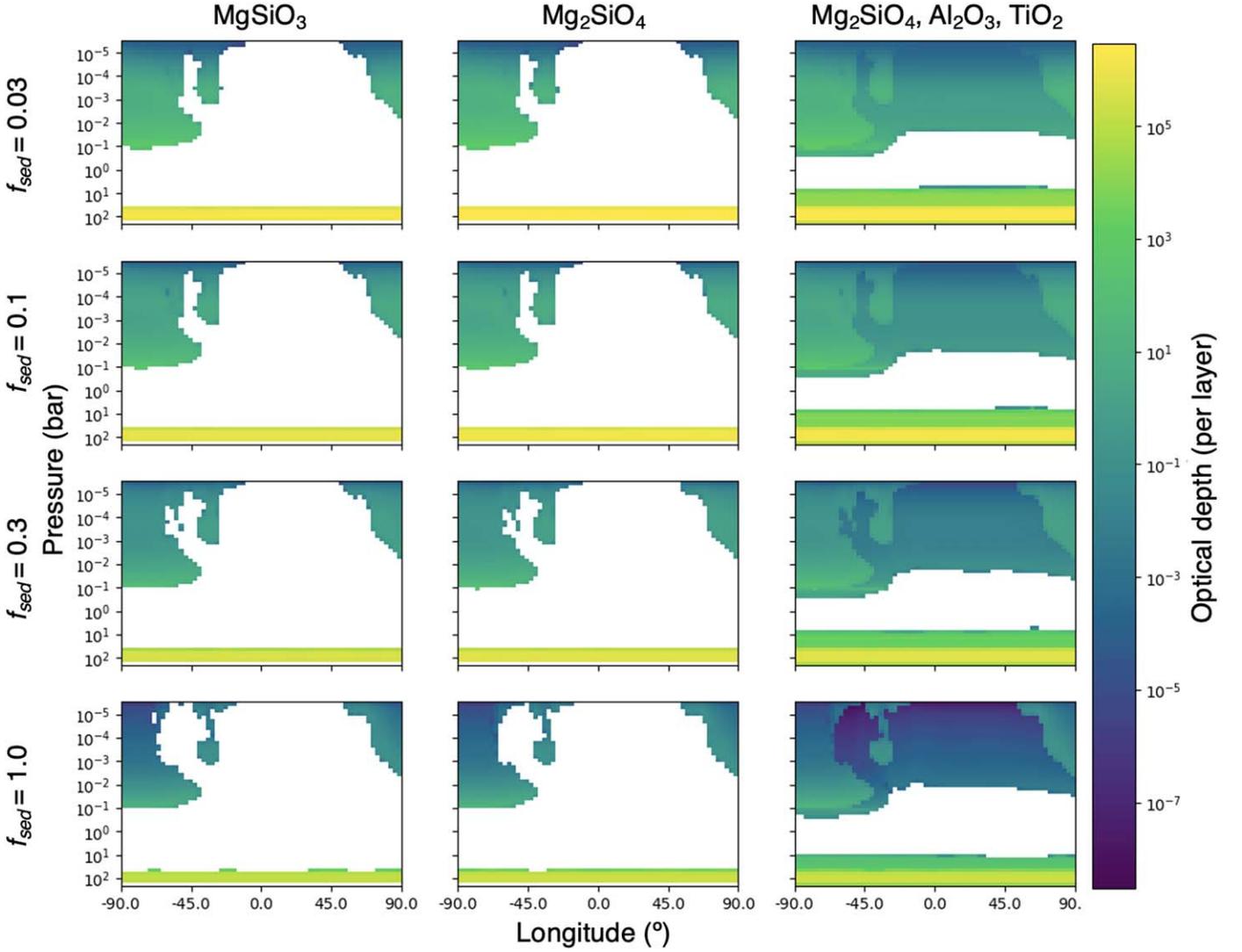

**Figure 4.** Cloud optical depth per layer for all cloudy scenarios as a function of longitude and pressure for the dayside hemisphere, at a latitude of 1°.4N and a wavelength of 604 nm. The columns represent the three cloud scenarios: $MgSiO_3$ only, $Mg_2SiO_4$ only, and $Mg_2SiO_4$, $Al_2O_3$, and $TiO_2$. The rows represent the four sedimentation efficiencies considered in this study (0.03, 0.1, 0.3, and 1.0). White regions indicate regions with no clouds.

Figure 5 shows the column optical depths for $Mg_2SiO_4$, $Al_2O_3$, and $TiO_2$ individually as a function of pressure for the four sedimentation efficiencies considered in this study. Increasing opacity of the plotted curves corresponds to decreasing values of $f_{sed}$. The blue curves show column optical depth at a grid point centered at (1°.4N, −68°.9W), which is near the western limb of the dayside hemisphere. The red curves show the column optical depth for a point centered at (1°.4N, 26°.7W), which is representative of the hot region east of the substellar point. The dichotomy of the blue and red curves in Figure 5 show how the differences in the thermal structures lead to drastically different cloud opacities at different locations on the dayside of Kepler-7b. For example, while the column optical depths of $Mg_2SiO_4$ clouds are greater than $10^{-4}$ for all $f_{sed}$ scenarios at the western limb, those on the hot dayside are zero for the entirety of the upper atmosphere (<30 bar). For $Al_2O_3$ and $TiO_2$, column optical depths are comparable in magnitude for both the western limb and the hot dayside, especially when $f_{sed}$ is low. For every $f_{sed}$ scenario shown, $Al_2O_3$ is the primary opacity source for the hot dayside region near the substellar point, while $Mg_2SiO_4$ is the dominant opacity source for the western limb.

### 3.2. Reflected-light Phase Curves for Kepler-7b

We compute theoretical reflected-light phase curves using PICASO by integrating spectroscopic relative-flux calculations across the dayside hemisphere visible to the observer. The presence of clouds can impact the spectroscopic relative flux, and thus the overall phase curve morphology and amplitude, depending on the cloud condensate species and chosen $f_{sed}$ values. Figure 6 shows examples of reflected-light albedo spectra as a planet-to-star flux ratio ($F_P/F_S$) in parts per million (ppm) for the entire dayside hemisphere at a phase of 0° for a variety of cloud condensate species with varying $f_{sed}$. We find that (1) the albedo of the cloudy scenarios increases with decreasing $f_{sed}$ values, (2) models with three cloud condensates ($Mg_2SiO_4$, $Al_2O_3$, and $TiO_2$) are more reflective than those that consider $Mg_2SiO_4$ or $MgSiO_3$ only, and (3) the $MgSiO_3$-only models are less reflective than the $Mg_2SiO_4$-only models. In general, the cloudless models result in much lower reflected-light intensity (0.6 ppm average) compared to the cloudy models. Most of the cloudless albedo is due to Rayleigh scattering, as seen by the slope from 400 to 600 nm.

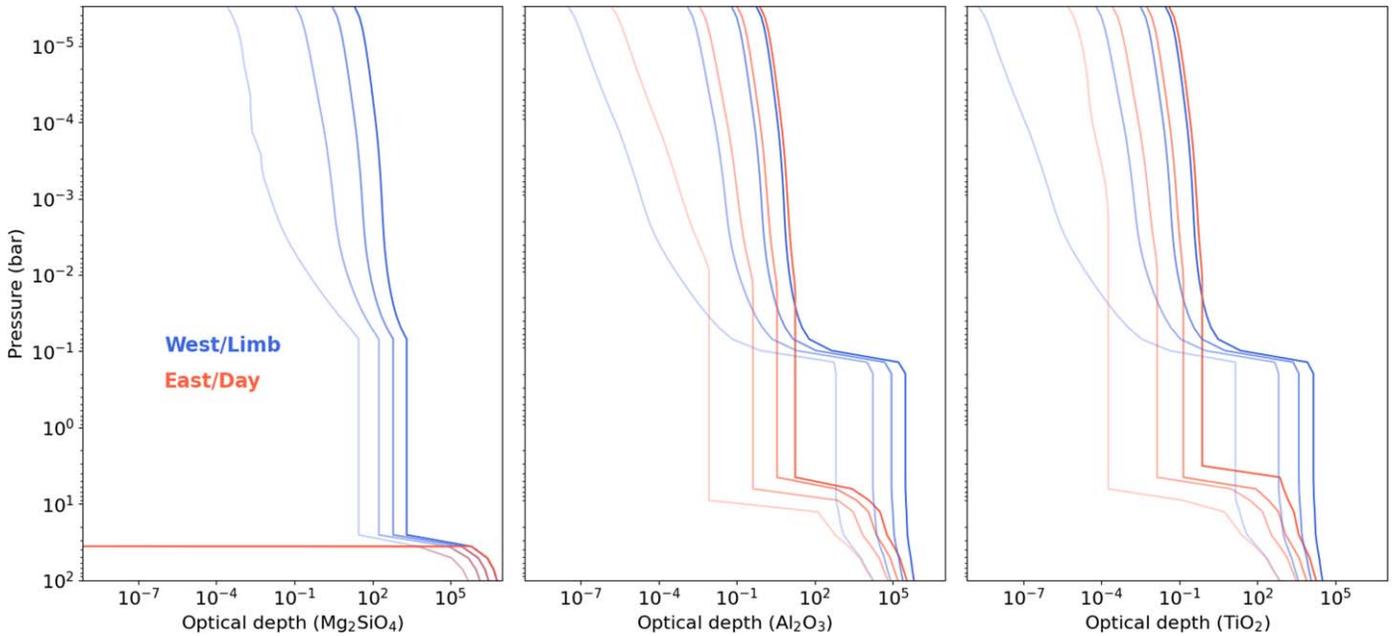

**Figure 5.** Wavelength-independent column optical depths for $Mg_2SiO_4$, $Al_2O_3$, and $TiO_2$, as a function of pressure, assuming geometric scatterers (see Equation (16) of A. S. Ackerman & M. S. Marley 2001). Blue corresponds to a western grid point centered at (1°4N, −68°9W), while red corresponds to an eastern grid point centered at (1°4N, 26°7W). The decreasing opacity of the plotted curves corresponds to increasing values of $f_{sed}$ (0.03, 0.1, 0.3, and 1.0).

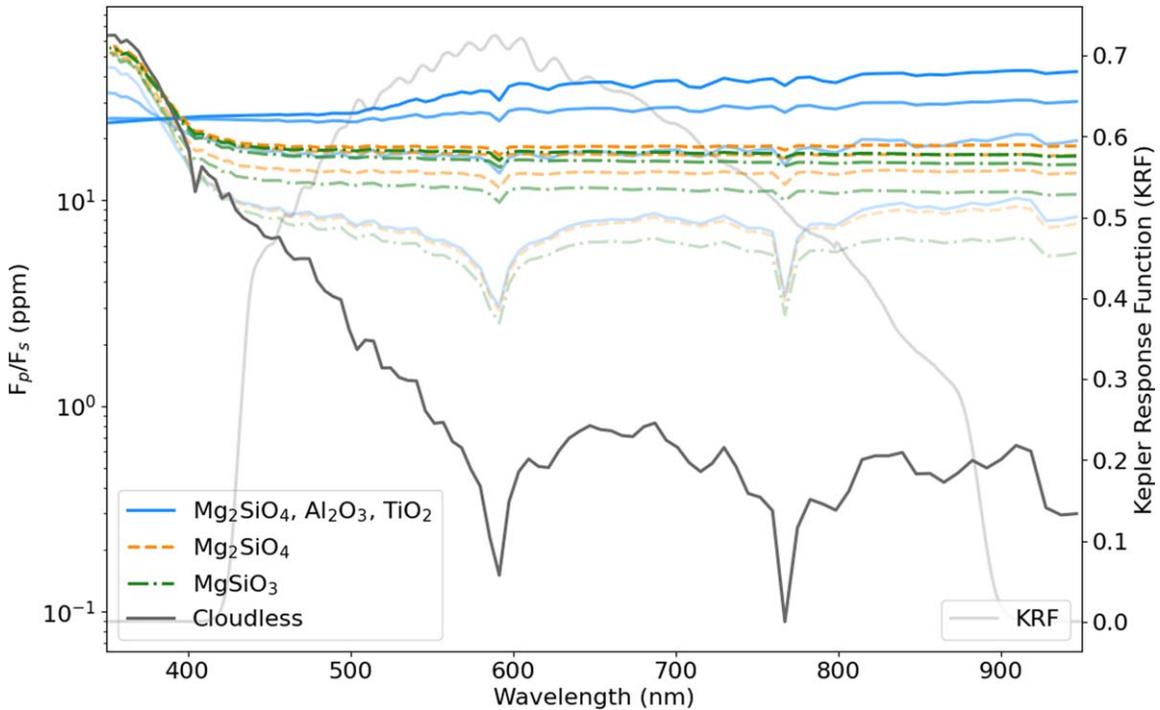

**Figure 6.** The planet-to-star flux ratio in parts per million (ppm) at phase 0° (secondary eclipse) for Kepler-7b for the cloudless case (black solid), and the $MgSiO_3$-only (green dashed–dotted lines), $Mg_2SiO_4$-only (orange dashed lines), and $Mg_2SiO_4$, $TiO_2$, and $Al_2O_3$ (blue solid lines) cloudy cases as calculated by `Virga`. The decreasing opacity of the curves corresponds to increasing values of $f_{sed}$ (0.03, 0.1, 0.3, and 1.0). The Kepler response function is plotted in light gray.

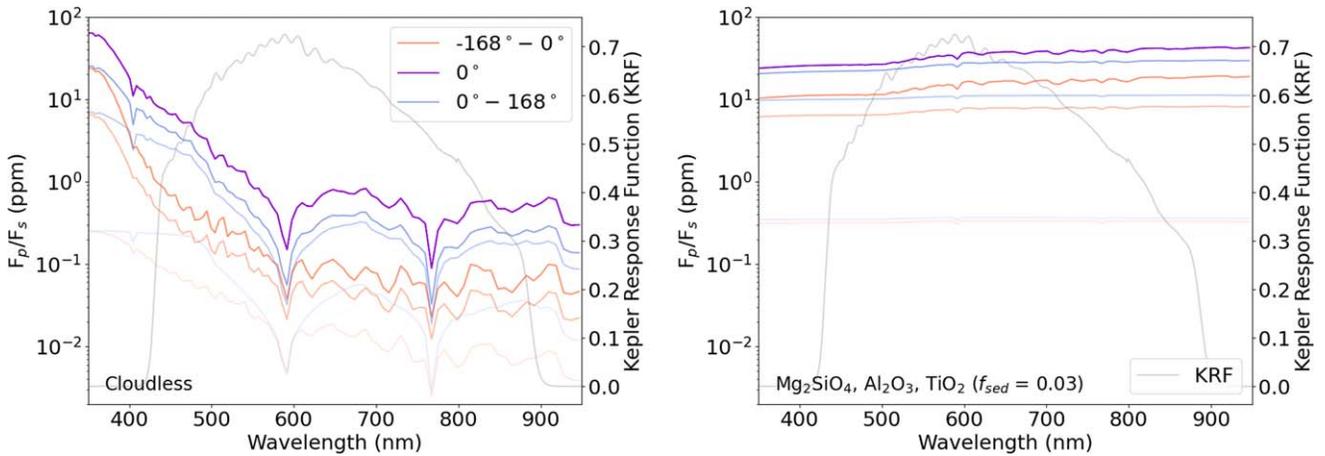

**Figure 7.** Albedo spectra at seven unique phase angles ranging from −168° to 168° for the cloudless scenario (left) and the cloudy scenario (right) in which $Mg_2SiO_4$, $Al_2O_3$, and $TiO_2$ are considered as cloud condensates with $f_{sed} = 0.03$. Red curves correspond to phases from −168° to 0°, and blue curves correspond to phases from 0° to 168°. A phase of 0° (secondary eclipse) is indicated by the purple curve. Increasing opacity of the curves corresponds to phases closer to 0°. The seven phase angles plotted are 0°, ±72°, ±120°, and ±168°. Phase angles of ±24°, ±48°, ±96°, ±144° are removed for plot clarity. The Kepler response function is plotted in gray.

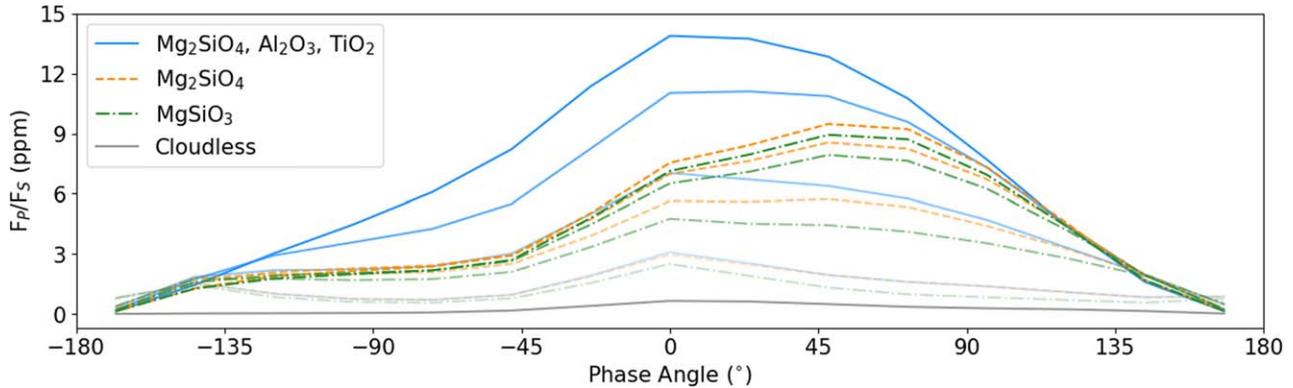

**Figure 8.** Reflected-light phase curves of Kepler-7b showing the planet-to-star flux ratio in parts per million (ppm) as a function of phase angle (−168° to 168°) for cloudless (gray solid), and $MgSiO_3$-only (green dashed–dotted), $Mg_2SiO_4$-only (orange dashed), and $Mg_2SiO_4$, $Al_2O_3$, and $TiO_2$ (blue solid) cloudy scenarios. The decreasing opacity of the curves corresponds to increasing values of $f_{sed}$ (0.03, 0.1, 0.3, and 1.0). The secondary eclipse occurs at 0° phase. Each phase curve has 15 unique phase angle computations and is weighed by the Kepler response function.

Figure 7 shows the relative flux changes for the cloudless scenario (left) and a cloudy scenario (right) with seven unique phase angles ranging from 0° to ±168°. The cloudy scenario is again showing the three-cloud regime ($Mg_2SiO_4$, $Al_2O_3$, and $TiO_2$) with $f_{sed} = 0.03$. As phase intensity approaches ±180°, the spectral intensity diminishes as less of the dayside hemisphere is visible to the observer. There are also differences in spectral intensity across phase angles for the cloudless case, due to the thermal/chemical differences across the dayside hemisphere. In general, the eastern (i.e., hotter) side exhibits more absorption features, due to the increase in molecular and atomic absorption cross sections with temperature. This results in more absorption features overall for negative phase angles where more of the eastern side is visible (red), compared to positive phase angles where more of the western side is visible (blue; Figure 7). For the cloudy scenario, there is a slight phase asymmetry in the spectral output across 0° phase due to the inhomogeneous cloud layer. The clouds on the western side of the substellar point (e.g., Figure 3) cause positive phase angles to exhibit higher albedos on average compared to negative phase angles.

The spectral outputs from PICASO are band-integrated and weighted by the KRF to produce the planet-to-star flux ratio phase curves in the Kepler band shown in Figure 8. The brightest phase curves come from the three-cloud scenarios ($Mg_2SiO_4$, $Al_2O_3$, and $TiO_2$) with the two lowest $f_{sed}$ values (0.03 and 0.1), with a peak phase curve amplitude of ∼11–14 ppm and a slight westward phase offset. The $Mg_2SiO_4$-only and $MgSiO_3$-only scenarios produce dimmer phase curves, especially at small and negative phase angles. The $Mg_2SiO_4$-only and $MgSiO_3$-only scenarios also produce phase curves with a much greater phase offset of ∼45°, with the exception of the highest $f_{sed}$ (1.0) scenario. As $f_{sed}$ increases, phase curve intensity decreases for all cloudy scenarios, due to the decreased vertical extent of the clouds.

The albedo differences between the three-cloud and $Mg_2SiO_4$-only models indicate that $Al_2O_3$ and $TiO_2$ can double the albedo of Kepler-7b, in particular near the secondary eclipse and at negative phase angles. This is because $Al_2O_3$ and $TiO_2$ nucleation occurs around the hot substellar point where no magnesium silicate nucleation can occur (Figure 4), thus increasing dayside albedo significantly. This finding suggests that, for hot Jupiters like Kepler-7b, which exhibit large variations in temperature across the dayside hemisphere, magnesium silicate clouds may not always be the dominant source of cloud opacity, as suggested by D. Powell et al. (2019) and P. Gao et al. (2020), especially for low-$f_{sed}$ scenarios. Our results also reinforce the previous finding that phase offsets are directly linked to cloud location and condensate species (V. Parmentier et al. 2016). In addition, the albedo differences between the MgSiO$_3$-only and Mg$_2$SiO$_4$-only models in Figure 8 are relatively minor (0–2 ppm), which is consistent with previous studies that found both magnesium silicate species can provide similar phase intensities when considering cloud particles produced through phase equilibrium (M. W. Webber et al. 2015).

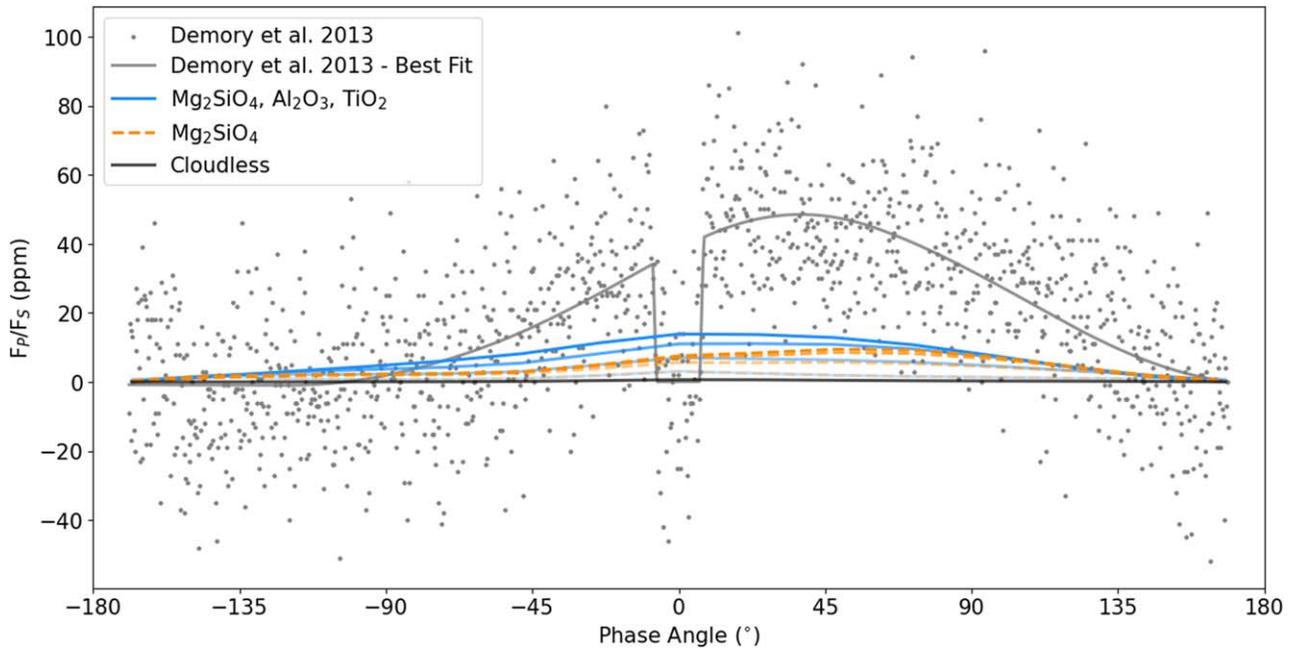

**Figure 9.** Same as Figure 8, but without the MgSiO$_3$-only scenario and including observational data (Q1–Q14 data from Kepler) of Kepler-7b from B.-O. Demory et al. (2013; gray points; error bars omitted for clarity). The gray line is the three-fixed-band best-fit model from B.-O. Demory et al. (2013).

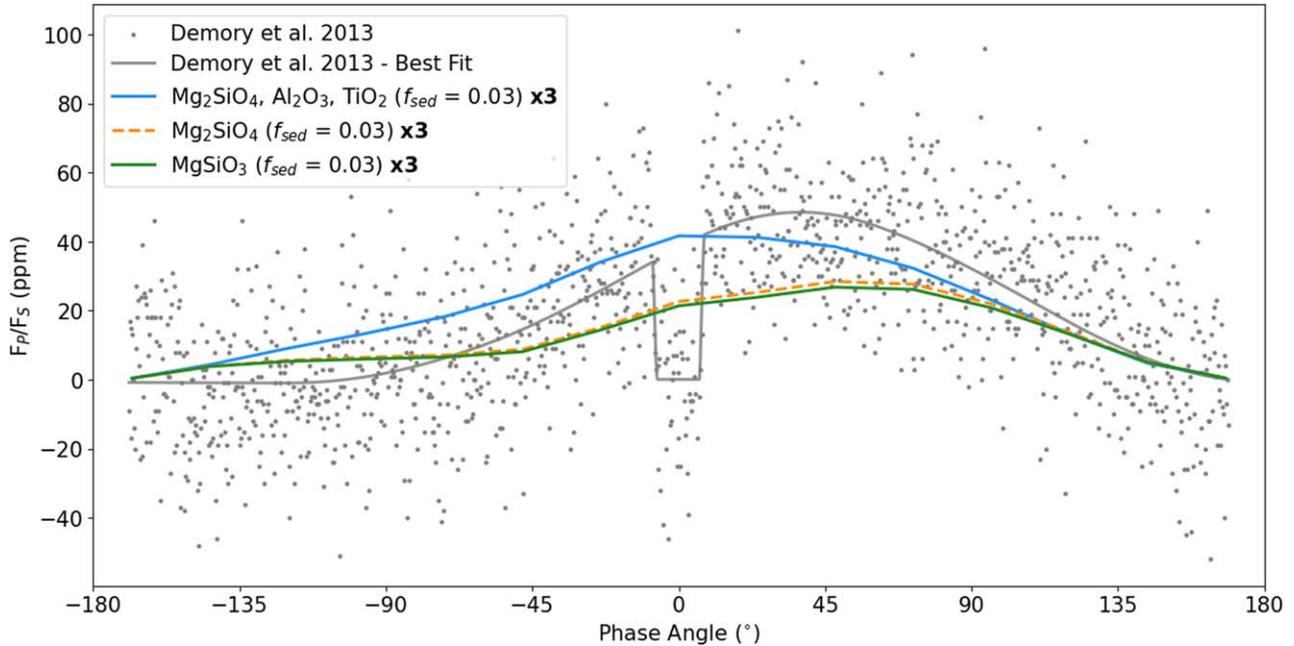

**Figure 10.** Same as Figure 9, but with only the $f_{sed} = 0.03$ models of the three-cloud (blue), Mg$_2$SiO$_4$ (orange), and MgSiO$_3$ (green) scenarios. The output intensities are multiplied by a factor of three in order to better compare the shapes of our reflected phase curves to the observations from B.-O. Demory et al. (2013; gray).

### 3.2.1. Comparison to Measurements

We compare our theoretical phase curves computed by PICASO to observations from the Kepler spacecraft originally published by B.-O. Demory et al. (2013), whose best-fitting models exhibit a phase curve maximum of $50 \pm 2$ ppm that is shifted toward positive phase angles by around 30° from the secondary eclipse (Figures 9 and 10). We choose to compare our model to B.-O. Demory et al. (2013) because their phase curve results are published as planet-to-star flux ratios, the same relative flux unit output by `PICASO`. Thus, our model can be compared to these observations without any further corrections. Other Kepler-7b phase curve observations (i.e., D. Angerhausen et al. 2015; L. J. Esteves et al. 2015) use a relative flux unit that includes Doppler and ellipsoidal fluctuations in the phase curve, which are not considered in our model. Despite these differences in data reduction, all three studies find a similar brightness amplitude of 48–50 ppm and a prominent westward phase shift.

Our output reflected-light phase curves for the three-cloud and $Mg_2SiO_4$ scenarios are shown in Figure 9 as compared to observational data and best-fit curve from B.-O. Demory et al. (2013). This best-fit curve assumes three discrete longitudinal brightness regions across the dayside hemisphere (i.e., three-fixed-band model). We underestimate the brightness amplitude by a factor of ~3 at secondary eclipse with the three-cloud scenario (Figure 10), which is similar to the findings of D. J. Adams et al. (2022). However, our low-$f_{sed}$ models match the shape of the observations well, especially for the $Mg_2SiO_4$ and $MgSiO_3$ scenarios as seen in Figure 10, where our output intensities are multiplied by a factor of three for ease of comparison. The three-cloud model has a larger degree of phase symmetry across secondary eclipse than both the $Mg_2SiO_4$/$MgSiO_3$ models and the best-fit curve from B.-O. Demory et al. (2013).

The discrepancies in phase intensity are due to our cloudy region west of the substellar point being too small and/or not bright enough. In addition, `Virga` does not consider zonal transport of cloud particles but instead considers each grid point individually when calculating cloud profiles. Zonal transport could substantially increase Kepler-7b's dayside brightness by transporting bright magnesium silicate particles from the nightside and western limb toward the substellar point, which would also produce a larger phase offset. Furthermore, the Kepler-7b GCM considered here does not treat effects from cloud radiative feedback. Opaque cloud formation may reduce dayside heating and increase cloud formation across the dayside hemisphere, thereby increasing the albedo of the planet (M. Roman & E. Rauscher 2017). Finally, `Virga` computes cloud scattering profiles for assumedly spherical particles. Realistic cloud particles are likely nonspherical and/or porous, which could lead to different particle scattering intensities and a larger vertical extent of particles in the atmosphere (M. G. Lodge et al. 2024; C. D. Hamill et al. 2024).

We employ a reduced chi-square test to quantify the goodness of fit between the observational data and our models. For our $\chi^2_{red}$ test, we ignore the observational data at $0° \pm 9°$ because our model does not simulate the actual secondary eclipse. For the $MgSiO_3$-only models, the $\chi^2_{red}$ values are 2.75, 2.50, 2.28, and 2.22, in order from highest to lowest sedimentation efficiency values. For the $Mg_2SiO_4$-only models, the $\chi^2_{red}$ values are 2.70, 2.42, 2.24, and 2.19. For the three-cloud models, the $\chi^2_{red}$ values are 2.70, 2.35, 2.07, and 1.94. High sedimentation efficiency values (0.3 and 1.0) result in the highest $\chi^2_{red}$ values and thus have the worst fit to observational data. The models with low sedimentation efficiency values (0.03 and 0.1) have the lowest $\chi^2_{red}$ values and provide a better fit. The single best-fit according to our $\chi^2_{red}$ test is the three-cloud scenario with a sedimentation efficiency of 0.03.

We also compared our model output to the closed-form ab initio model by K. Heng et al. (2021) used to calculate the geometric albedo and reflected-light phase curve for Kepler-7b. K. Heng et al. (2021) calculated a best-fit-model to Kepler-7b phase curve observations with the following parameters: a cloudless region from $-6°$ to $40°$ longitude, with $0°$ as the substellar point, where the cloudless region has a single-scattering albedo of 0.0136. All other regions are considered cloudy with a single-scattering albedo of 0.1286. The single-scattering albedos output by `Virga` are ~0.95 for the cloudy regions on the western dayside and eastern terminator, and ~0.7–0.8 for the dim region around the substellar point. Thus, the parameters from K. Heng et al. (2021) will yield much lower albedos than our models when used in `PICASO`, likely due to the spectral sensitivity of `PICASO` that is lost when generalizing albedo across all wavelengths and atmospheric heights.

Despite the assumptions inherent in the GCM and `Virga`, and the complications involved in recreating Kepler-7b's bright albedo (e.g., M. W. Webber et al. 2015; D. J Adams et al. 2022), these results show that `PICASO` can provide a useful tool for comparing phase-resolved models to observations. Coupled with its ability to produce thermal emission phase curves (N. Robbins-Blanch et al. 2022), `PICASO` can provide a better understanding of the albedo and atmospheric properties of exoplanet atmospheres observed at visible and near-infrared wavelengths.

## 4. Conclusions

In this paper, we present an updated routine for computing phase-resolved reflected-light observations with and without clouds in `PICASO` using 3D GCMs as input. Our updates are applied to the original phase curve formulation created for thermal emission by N. Robbins-Blanch et al. (2022), and we integrate the previous work of D. J. Adams et al. (2022), which allows `PICASO` to compute reflected-light spectra for a 3D atmosphere. We take into account the changing illuminated portion of the planet visible to the observer by rotating the GCM and cloud distributions within the model. We apply our new technique to Kepler-7b, a hot Jupiter with a westward shifted phase curve maximum in visible light likely caused by nonuniform, reflective cloud layers. We use the thermal and vertical mixing profiles from a SPARC/MITgcm of Kepler-7b as input to `PICASO`. We calculate cloud location and vertical extent using `Virga` assuming only $MgSiO_3$ clouds, only $Mg_2SiO_4$ clouds, and a three-cloud regime ($Mg_2SiO_4$, $Al_2O_3$, and $TiO_2$). Each scenario was modeled with $f_{sed}$ values of 0.03, 0.1, 0.3, and 1.0, for a total of 12 cloud models. In addition, we consider a cloudless atmosphere.

While our three-cloud models match the general shape of the Kepler phase curve observations, it does not reproduce the observed phase curve amplitude or phase offset stemming from the cloudy western side. This is likely because `Virga` and `PICASO` underestimate the true extent and brightness of the western clouds, since these models do not take zonal transport, cloud radiative feedback, and particle nonsphericity into account. Our best-fitting models to the observational data suggest that Kepler-7b is home to multiple cloud species, $Mg_2SiO_4$, $Al_2O_3$, and $TiO_2$, with low (0.03, 0.1) $f_{sed}$ values. Our results suggest $Al_2O_3$ and $TiO_2$ may contribute up to half of Kepler-7b's albedo, which contradicts previous studies that focused primarily on transmission spectroscopy, which predict magnesium silicate clouds to be the dominant opacity source for hot Jupiters with an equilibrium temperature ∼1500–1600 K (D. Powell et al. 2019; P. Gao et al. 2020).

Reflected-light phase curve models will become increasingly important as the next generation of space telescopes takes flight. Observations with the Near-Infrared Imager and Slitless Spectrograph/Single-Object Slitless Spectroscopy (L. Albert et al. 2023) aboard JWST have already begun to observe the scattering signatures of cloudy exoplanets (A. D. Feinstein et al. 2023). Both the Nancy Grace Roman Space Telescope and the Habitable Worlds Observatory will have the ability to probe the reflected-light signal of exoplanets in the visible and near-infrared (N. J. Kasdin (Kasdin et al. 2020; S. R. Vaughan et al. 2023). Having a prior understanding of the brightness of these exoplanets under a variety of conditions will help us decipher future observations of these distant worlds.

## Acknowledgments

We acknowledge support from the NASA Exoplanet Research Program (80NSSC23K0041). We thank Brice-Olivier Demory for providing us with Kepler data and the best-fit curve. D.J.A. was supported by NASA through the NASA Hubble Fellowship grant #HST-HF2-51523.001-A awarded by the Space Telescope Science Institute, which is operated by the Association of Universities for Research in Astronomy, Inc., for NASA, under contract NAS5-26555.

*Software*: PICASO (N. Batalha et al. 2024), virga (N. Batalha et al. 2020b), Matplotlib (J. D. Hunter 2007), pickle (G. Van Rossum 2020), bokeh (Bokeh Development Team 2014), NumPy (S. van der Walt et al. 2011).

## Appendix
## Spatial Resolution Sensitivity Tests

We tested the sensitivity of PICASO's reflected-light intensity output with respect to the spatial resolution of the dayside hemisphere in order to find the optimal balance between precision and computation time. We compare the intensity output of the cloudy model that considers $Mg_2SiO_4$, $Al_2O_3$, and $TiO_2$ as cloud condensates with $f_{sed} = 0.03$ at three different spatial resolutions: 10 × 10, 20 × 20, and 30 × 30 (Gauss angles × Chebyshev angles) in Figure 11. We find that a 10 × 10 grid produces percentage errors greater than 5% when compared to 20 × 20 and 30 × 30 grid models at select phase angles. While the 10 × 10 models are much quicker to compute compared to the higher-resolution models, we did not find this error acceptable for this study. The 20 × 20 model, however, produces errors that do not surpass 2% when compared to the 30 × 30 case. We thus find that the 20 × 20 model offers the best balance between computation time and precision. The 20 × 20 and 30 × 30 models were calculated with a high-performance computer (12-core central processing unit, 96 GB of RAM). With this computer, the 10 × 10 phase curves took several minutes to compute, the 20 × 20 phase curves took one to two hours to compute, and the 30 × 30 models took nine hours to two days to compute.

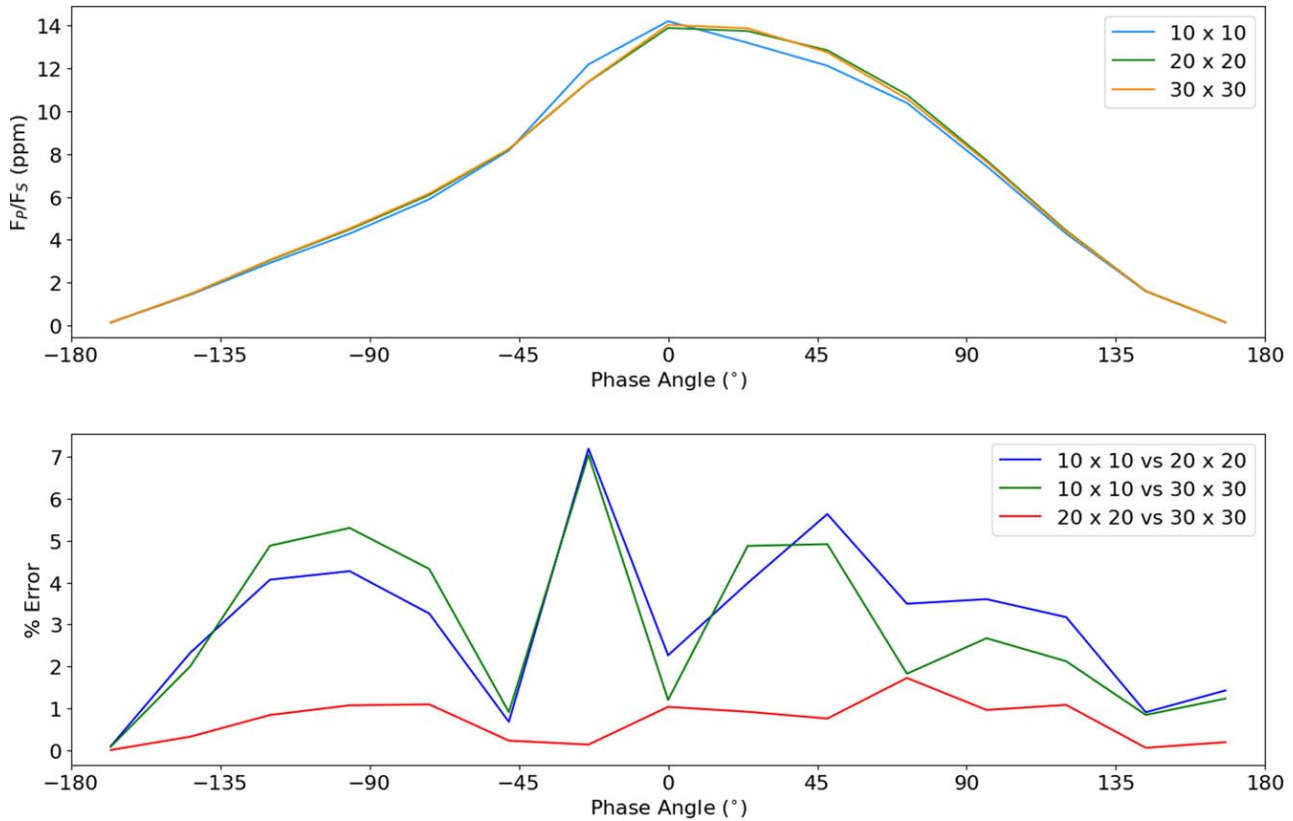

**Figure 11.** Top: comparison of the phase curves for the three-cloud scenario that considers $Mg_2SiO_4$, $Al_2O_3$, and $TiO_2$ as cloud condensates with $f_{sed} = 0.03$ for three different Chebyshev–Gauss spatial resolutions: 10 × 10 (blue), 20 × 20 (green), and 30 × 30 (orange). Bottom: absolute percent error between each of the models as a function of orbital phase.

## ORCID iDs

Colin D. Hamill https://orcid.org/0000-0002-9464-8494
Alexandria V. Johnson https://orcid.org/0000-0002-6227-3835
Natasha Batalha https://orcid.org/0000-0003-1240-6844
Rowan Nag https://orcid.org/0009-0003-0371-296X
Peter Gao https://orcid.org/0000-0002-8518-9601
Danica Adams https://orcid.org/0000-0001-9897-9680
Tiffany Kataria https://orcid.org/0000-0003-3759-9080

## References

Abramowitz, M., & Stegun, I. A. 1972, Handbook of Mathematical Functions (New York: Dover)
Ackerman, A. S., & Marley, M. S. 2001, ApJ, 556, 872
Adams, D. J., Kataria, T., Batalha, N. E., Gao, P., & Knutson, H. A. 2022, ApJ, 926, 157
Albert, L., Lafrenière, D., Doyon, R., et al. 2023, PASP, 135, 075001
Allard, F., Allard, N. F., Homeier, D., et al. 2007, A&A, 474, L21
Allard, N. F., Kielkopf, J. F., & Allard, F. 2007, EPJD, 44, 507
Allard, N. F., Spiegelman, F., & Kielkopf, J. F. 2016, A&A, 589, A21
Allard, N. F., Spiegelman, F., Leininger, T., & Molliere, P. 2019, A&A, 628, A120
Amundsen, D. S., Mayne, N. J., Baraffe, I., et al. 2016, A&A, 595, A36
Angerhausen, D., DeLarme, E., & Morse, J. A. 2015, PASP, 127, 1113
Arcangeli, J., Désert, J.-M., Parmentier, V., et al. 2019, A&A, 625, A136
Batalha, N. 2020, natashabatalha/virga: Initial Release v0.0, Zenodo, doi:10.5281/zenodo.3759888
Batalha, N., Freedman, R., Lupu, R., & Marley, M. 2020a, Resampled Opacity Database for PICASO v2, Zenodo, doi:10.5281/zenodo.3759675
Batalha, N., & Marley, M. 2020, Refractive Indices for Virga Exoplanet Cloud Model v1.1, Zenodo, doi:10.5281/zenodo.3992294
Batalha, N. E., Marley, M. S., Lewis, N. K., & Fortney, J. J. 2019, ApJ, 878, 70
Batalha, N., & Rooney, C. 2020, natashabatalha/picaso: Release v2.1, Zenodo, doi:10.5281/zenodo.4206648
Batalha, N., Rooney, C., & Mukherjee, S. 2020b, natashabatalha/virga: Initial Release, v0.0, Zenodo, doi:10.5281/zenodo.3759888
Batalha, N., Rooney, C., Mukherjee, S., et al. 2024, natashabatalha/picaso: Release 2.3, v3.3, Zenodo, doi:10.5281/zenodo.14160128
Baudino, J.-L., Bézard, B., Boccaletti, A., et al. 2015, A&A, 582, A83
Baudino, J.-L., Mollière, P., Venot, O., et al. 2017, ApJ, 850, 150
Bean, J. L., Kempton, E. M.-R., & Homeier, D. 2010, Natur, 468, 669
Beatty, T. G., Wong, I., Fetherolf, T., et al. 2020, AJ, 160, 211
Bokeh Development Team 2014, Bokeh: Python library for interactive visualization, http://www.bokeh.pydata.org
Bruno, G., Lewis, N. K., Stevenson, K. B., et al. 2018, AJ, 155, 55
Cahoy, K. L., Marley, M. S., & Fortney, J. J. 2010, ApJ, 724, 189
Changeat, Q., Skinner, J. W., Cho, J. Y.-K., et al. 2024, ApJS, 270, 34
Charbonneau, D., Brown, T. M., Noyes, R. W., & Gilliland, R. L. 2002, ApJ, 568, 377
Charnay, B., Bézard, B., Baudino, J.-L., et al. 2018, ApJ, 854, 172
Chubb, K. L., Stam, D. M., Helling, C., Samra, D., & Carone, L. 2024, MNRAS, 527, 4955
Christie, D. A., Mayne, N. J., Lines, S., et al. 2021, MNRAS, 506, 4500
Coughlin, J. L., & López-Morales, M. 2012, AJ, 143, 39
Crossfield, I. J.-M. 2015, PASP, 127, 941
Daylan, T., Günther, M. N., Mikal-Evans, T., et al. 2021, AJ, 161, 131
Demory, B.-O., de Wit, J., Lewis, N., et al. 2013, ApJL, 776, L25
Demory, B.-O., Seager, S., Madhusudhan, N., et al. 2011, ApJL, 735, L12
Dobbs-Dixon, I., & Agol, E. 2013, MNRAS, 435, 3159
Dyrek, A., Min, M., Decin, L., et al. 2024, Natur, 625, 51
Esteves, L. J., Mooij, E. J.-W. D., & Jayawardhana, R. 2013, ApJ, 772, 51
Esteves, L. J., Mooij, E. J.-W. D., & Jayawardhana, R. 2015, ApJ, 804, 150
Feinstein, A. D., Radica, M., Welbanks, L., et al. 2023, Natur, 614, 670
Flowers, E., Brogi, M., Rauscher, E., Kempton, E. M.-R., & Chiavassa, A. 2019, AJ, 157, 209
Fortney, J. J., Cooper, C. S., Showman, A. P., Marley, M. S., & Freedman, R. S. 2006, ApJ, 652, 746
Fortney, J. J., Lodders, K., Marley, M. S., & Freedman, R. S. 2008, ApJ, 678, 1419
Fortney, J. J., Marley, M. S., Lodders, K., Saumon, D., & Freedman, R. 2005, ApJL, 627, L69
Fortney, J. J., Shabram, M., Showman, A. P., et al. 2010, ApJ, 709, 1396
Fraine, J., Mayorga, L. C., Stevenson, K. B., et al. 2021, AJ, 161, 269
Gao, P., Marley, M. S., & Ackerman, A. S. 2018, ApJ, 855, 86
Gao, P., Piette, A. A. A., Steinrueck, M. E., et al. 2023, ApJ, 951, 96
Gao, P., Thorngren, D. P., Lee, E. K.-H., et al. 2020, NatAs, 4, 951
Gao, P., Wakeford, H. R., Moran, S. E., & Parmentier, V. 2021, JGRE, 126, e2020JE006655
Gibson, N. P., Nikolov, N., Sing, D. K., et al. 2017, MNRAS, 467, 4591
Goody, R. M., & Yung, Y. L. 1989, Atmospheric Radiation: Theoretical Basis (Oxford: Oxford Univ. Press)
Gordon, S., & Mcbride, B. J. 1994, Computer program for calculation of complex chemical equilibrium compositions and applications. Part 1: Analysis, NTRS, https://ntrs.nasa.gov/citations/19950013764
Grant, D., Lewis, N. K., Wakeford, H. R., et al. 2023, ApJL, 956, L29
Hamill, C. D., Johnson, A. V., & Gao, P. 2024, PSJ, 5, 186
Harada, C. K., Kempton, E. M.-R., Rauscher, E., et al. 2021, ApJ, 909, 85
Helling, C. 2019, AREPS, 47, 583
Helling, C., Ackerman, A., Allard, F., et al. 2008, MNRAS, 391, 1854
Helling, C., & Fomins, A. 2013, RSPTA, 371, 20110581
Heng, K., & Demory, B.-O. 2013, ApJ, 777, 100
Heng, K., Morris, B. M., & Kitzmann, D. 2021, NatAs, 5, 1001
Hu, R., Demory, B.-O., Seager, S., Lewis, N., & Showman, A. P. 2015, ApJ, 802, 51
Hunter, J. D. 2007, CSE, 9, 3
Jackson, B., Adams, E., Sandidge, W., Kreyche, S., & Briggs, J. 2019, AJ, 157, 239
Jansen, T., & Kipping, D. 2020, MNRAS, 494, 4077
Kasdin, N. J., Bailey, V. P., Mennesson, B., et al. 2020, Proc. SPIE, 11443, 300
Kataria, T., Showman, A. P., Fortney, J. J., Marley, M. S., & Freedman, R. S. 2014, ApJ, 785, 92
Kataria, T., Showman, A. P., Fortney, J. J., Marley, M. S., & Freedman, R. S. 2015, ApJ, 801, 86
Kataria, T., Showman, A. P., Lewis, N. K., et al. 2013, ApJ, 767, 76
Kataria, T., Sing, D. K., Lewis, N. K., et al. 2016, ApJ, 821, 9
Kawashima, Y., & Ikoma, M. 2018, ApJ, 853, 7
Kempton, E. M.-R., Zhang, M., Bean, J. L., et al. 2023, Natur, 620, 67
Kipping, D., & Bakos, G. 2011, ApJ, 730, 50
Knutson, H. A., Benneke, B., Deming, D., & Homeier, D. 2014, Natur, 505, 66
Knutson, H. A., Charbonneau, D., Allen, L. E., et al. 2007, Natur, 447, 183
Kreidberg, L., Bean, J. L., Désert, J.-M., et al. 2014, Natur, 505, 69
Kreidberg, L., Line, M. R., Parmentier, V., et al. 2018, AJ, 156, 17
Latham, D. W., Borucki, W. J., Koch, D. G., et al. 2010, ApJL, 713, L140
Lewis, N. K., Parmentier, V., Kataria, T., et al. 2017, arXiv:1706.00466
Lewis, N. K., Showman, A. P., Fortney, J. J., Knutson, H. A., & Marley, M. S. 2014, ApJ, 795, 150
Lodge, M. G., Wakeford, H. R., & Leinhardt, Z. M. 2024, MNRAS, 527, 11113
Lustig-Yaeger, J., Fu, G., May, E. M., et al. 2023, NatAs, 7, 1317
Ma, S., Ito, Y., Al-Refaie, A. F., et al. 2023, ApJ, 957, 104
MacDonald, R. J., Marley, M. S., Fortney, J. J., & Lewis, N. K. 2018, ApJ, 858, 69
Marley, M. S, Ackerman, A. S., Cuzzi, J. N., & Kitzmann, D. 2013, in Comparative Climatology of Terrestrial Planets, ed. S. J. Mackwell et al. (Tucson, AZ: Univ. of Arizona Press), 367
Marley, M. S., Gelino, C., Stephens, D., Lunine, J. I., & Freedman, R. 1999, ApJ, 513, 879
Marley, M. S., & McKay, C. P. 1999, Icar, 138, 268
Marley, M. S., Saumon, D., & Goldblatt, C. 2010, ApJL, 723, L117
Marley, M. S., Saumon, D., Visscher, C., et al. 2021, ApJ, 920, 85
May, E. M., Komacek, T. D., Stevenson, K. B., et al. 2021, AJ, 162, 158
McKay, C. P., Pollack, J. B., & Courtin, R. 1989, Icar, 80, 23
Mendonça, J. M., Malik, M., Demory, B.-O., & Heng, K. 2018, AJ, 155, 150
Morello, G., Changeat, Q., Dyrek, A., Lagage, P.-O., & Tan, J. C. 2023, A&A, 676, A54
Morley, C. V., Fortney, J. J., Kempton, E. M.-R., et al. 2013, ApJ, 775, 33
Morley, C. V., Fortney, J. J., Marley, M. S., et al. 2012, ApJ, 756, 172
Morley, C. V., Fortney, J. J., Marley, M. S., et al. 2015, ApJ, 815, 110
Morley, C. V., Marley, M. S., Fortney, J. J., et al. 2014, ApJ, 787, 78
Mukherjee, S., Batalha, N. E., Fortney, J. J., & Marley, M, S. 2023, ApJ, 942, 71
Mukherjee, S., Fortney, J. J., Batalha, N. E., et al. 2022, ApJ, 938, 107
Mukherjee, S., Fortney, J. J., Morley, C. V., et al. 2024, ApJ, 963, 73
Muñoz, A. G., & Isaak, K. G. 2015, PNAS, 112, 13461
Ohno, K., Okuzumi, S., & Tazaki, R. 2020, ApJ, 891, 131
Oreshenko, M., Heng, K., & Demory, B.-O. 2016, MNRAS, 457, 3420
Parmentier, V., Fortney, J. J., Showman, A. P., Morley, C., & Marley, M. S. 2016, ApJ, 828, 22
Parmentier, V., Line, M. R., Bean, J. L., et al. 2018, A&A, 617, A110

Parmentier, V., Showman, A. P., & Fortney, J. J. 2021, MNRAS, 501, 78
Parmentier, V., Showman, A. P., & Lian, Y. 2013, A&A, 558, A91
Polyansky, O. L., Kyuberis, A. A., Zobov, N. F., et al. 2018, MNRAS, 480, 2597
Powell, D., Louden, T., Kreidberg, L., et al. 2019, ApJ, 887, 170
Quintana, E. V., Rowe, J. F., Barclay, T., et al. 2013, ApJ, 767, 137
Robbins-Blanch, N., Kataria, T., Batalha, N. E., & Adams, D. J. 2022, ApJ, 930, 93
Roman, M., & Rauscher, E. 2017, ApJ, 850, 17
Roman, M. T., Kempton, E. M.-R., Rauscher, E., et al. 2021, ApJ, 908, 101
Rooney, C. M., Batalha, N. E., Gao, P., & Marley, M. S. 2022, ApJ, 925, 33
Ryabchikova, T., Piskunov, N., Kurucz, R. L., et al. 2015, PhyS, 90, 054005
Samra, D., Helling, Ch., & Min, M. 2020, A&A, 639, A107
Samra, D., Helling, Ch., & Birnstiel, T. 2022, A&A, 663, A47
Saumon, D., & Marley, M. S. 2008, ApJ, 689, 1327
Shporer, A., & Hu, R. 2015, AJ, 150, 112
Shporer, A., O'Rourke, J. G., Knutson, H. A., et al. 2014, ApJ, 788, 92
Shporer, A., Wong, I., Huang, C. X., et al. 2019, AJ, 157, 178
Sing, D. K., Fortney, J. J., Nikolov, N., et al. 2016, Natur, 529, 59
Skemer, A. J., Morley, C. V., Zimmerman, N. T., et al. 2016, ApJ, 817, 166
Skinner, J. W., & Cho, J. Y.-K. 2022, MNRAS, 511, 3584
Steinrueck, M. E., Parmentier, V., Showman, A. P., Lothringer, J. D., & Lupu, R. E. 2019, ApJ, 880, 14
Stevenson, K. B., Désert, J.-M., Line, M. R., et al. 2014, Sci, 346, 838
Stevenson, K. B., Line, M. R., Bean, J. L., et al. 2017, AJ, 153, 68
Tang, S.-Y., Robinson, T. D., Marley, M. S., et al. 2021, ApJ, 922, 26
Tazaki, R., & Tanaka, H. 2018, ApJ, 860, 79
Thao, P. C., Mann, A. W., Johnson, M. C., et al. 2020, AJ, 159, 32
Toon, O. B., McKay, C. P., Ackerman, T. P., & Santhanam, K. 1989, JGR, 94, 16287
van der Walt, S., Colbert, S. C., & Varoquaux, G. 2011, CSE, 13, 22
Van Rossum, G. 2020, The Python Library Reference v3.8.2, Python, https://www.python.org/downloads/release/python-382/
Vaughan, S. R., Gebhard, T. D., Bott, K., et al. 2023, MNRAS, 524, 5477
Venot, O., Parmentier, V., Blecic, J., et al. 2020, ApJ, 890, 176
Visscher, C., Lodders, K., & Fegley, B. 2010, ApJ, 716, 1060
Visscher, C., & Moses, J. I. 2011, ApJ, 738, 72
von Essen, C., Mallonn, M., Borre, C. C., et al. 2020, A&A, 639, A34
Wakeford, H. R., Sing, D. K., Kataria, T., et al. 2017, Sci, 356, 628
Webber, M. W., Lewis, N. K., Marley, M., et al. 2015, arXiv:1503.01028
Windsor, J. D., Robinson, T. D., Kopparapu, R. K., et al. 2023, PSJ, 4, 94
Wong, I., Benneke, B., Shporer, A., et al. 2020, AJ, 159, 104
Wong, I., Kitzmann, D., Shporer, A., et al. 2021, AJ, 162, 127
Wong, I., Knutson, H. A., Kataria, T., et al. 2016, ApJ, 823, 122
Zellem, R. T., Lewis, N. K., Knutson, H. A., et al. 2014, ApJ, 790, 53
Zhang, X., & Showman, A. P. 2018, ApJ, 866, 2